\documentclass[%
 reprint,
 amsmath,amssymb,
 aps,
prb,
floatfix,
showkeys
]{revtex4-2}
\usepackage{xcolor}
\usepackage{graphicx}
\usepackage{dcolumn}
\usepackage{bm}
\usepackage{hyperref}
\usepackage{ulem}

\hypersetup{
  colorlinks   = true, 
  urlcolor     = green, 
  linkcolor    = blue, 
  citecolor   = blue 
}

\begin{document}

\preprint{?}

\title{Anomalous microwave response in the dissipative regime of topological superconducting devices based on Bi$_2$Te$_{2.3}$Se$_{0.7}$}

\author{Vasily Stolyarov}
\affiliation{%
 Advanced Mesoscience and Nanotechnology Centre, Moscow Institute of Physics and Technology,
141700 Dolgoprudny, Russia
}

\author{Sergei Kozlov}
\affiliation{%
 Laboratoire de Physique et d'étude des Matériaux, ESPCI Paris, CNRS, PSL University, 75005 Paris, France
}

\author{Dmitry Yakovlev}
\email{dimitry.yakovlev@espci.fr}
\affiliation{%
 Laboratoire de Physique et d'étude des Matériaux, ESPCI Paris, CNRS, PSL University, 75005 Paris, France
}

\author{Nicolas Bergeal}
\affiliation{%
 Laboratoire de Physique et d'étude des Matériaux, ESPCI Paris, CNRS, PSL University, 75005 Paris, France
}%

\author{Cheryl Feuillet-Palma}
\affiliation{%
 Laboratoire de Physique et d'étude des Matériaux, ESPCI Paris, CNRS, PSL University, 75005 Paris, France
}%

\author{Olga Skryabina}
\affiliation{%
 Advanced Mesoscience and Nanotechnology Centre, Moscow Institute of Physics and Technology,
141700 Dolgoprudny, Russia
}%

\author{Dmitry Lvov}
\affiliation{%
 Department of Applied Physics, Low Temperature Laboratory, Aalto University School of Science, PO Box 13500, 00076 Aalto, Finland
}%

\author{Mikhail Kupriyanov}
\affiliation{%
 Skobeltsyn Institute of Nuclear Physics, Lomonosov Moscow State University, 119991 Moscow, Russia
}%

\author{Alexander Golubov}
\affiliation{%
 Faculty of Science and Technology and MESA$^+$ Institute for Nanotechnology, University of Twente, 7500 AE Enschede, The Netherlands;
}%

\author{Dimitri Roditchev}
\affiliation{%
 Laboratoire de Physique et d'étude des Matériaux, ESPCI Paris, CNRS, PSL University, 75005 Paris, France
}%

\date{\today}

\begin{abstract}
Superconducting proximity junctions based on topological insulators are widely believed to harbor Majorana-like bound states. The latter serves as a paradigm non-local topological quantum computation protocols. Nowadays, a search for topological phases in different materials, perspective for a realization of topological qubits, is one of the central efforts in quantum physics. It is motivated, in particular, by recent observation of anomalous ac Josephson effect, which being a signature of Majorana physics. Its manifestations, such as a fractional Josephson frequency and the absence of the first (or several odd in more rare cases), Shapiro steps, were reported for different materials.

Here we study Shapiro steps in Nb/Bi$_2$Te$_{2.3}$Se$_{0.7}$/Nb junctions, based on ultrasmall single crystals of a 3D topological insulator synthesized by a physical vapor deposition (PVD) technique. We present evidence that our junctions are ballistic. When subjected to microwave radiation, the junctions exhibit Shapiro steps, but the first step is missing. Typically it is assumed that the missing first step (MFS) effect cannot be observed in the presence of quasiparticle poisoning due to suppression of the 4$\pi$-periodic component. Our findings within the context of the RSJ-model of Josephson junction dynamics show that such behaviour of samples corresponds to a specific condition, requiring a minimum of 5\% of the 4$\pi$-component for disappearance of the first Shapiro step.
\end{abstract}


\maketitle

\section{\label{sec:Introduction}Introduction}

Andreev Bound States (ABS) emerge in Superconductor/Normal metal/Superconductor (SNS) Josephson junctions as localized solutions of the Bogoliubov-De Gennes equations within the normal region \cite{sauls2018andreev}. In SNS junctions with topological order in the normal part, Majorana Zero Modes (MZM) are believed to appear ~\cite{rokhinson2012fractional, mourik2012signatures, das2012zero, houzet2013dynamics, jiang2011unconventional}. Localized Majorana fermions encode topological Andreev Bound States ($T$-ABS) which coherently transfer $1e$ charge and, correspondingly, their current-phase relationship (CPR) is $4\pi$-periodic, at least according to the theory. Such periodicity is possible since the topological superconductors can accept single electrons onto a pair of Majorana modes, and such single electrons have zero energy, unlike the usual quasiparticles (bogoliubons).

\begin{figure}[t!]
\includegraphics[width= 8 cm]{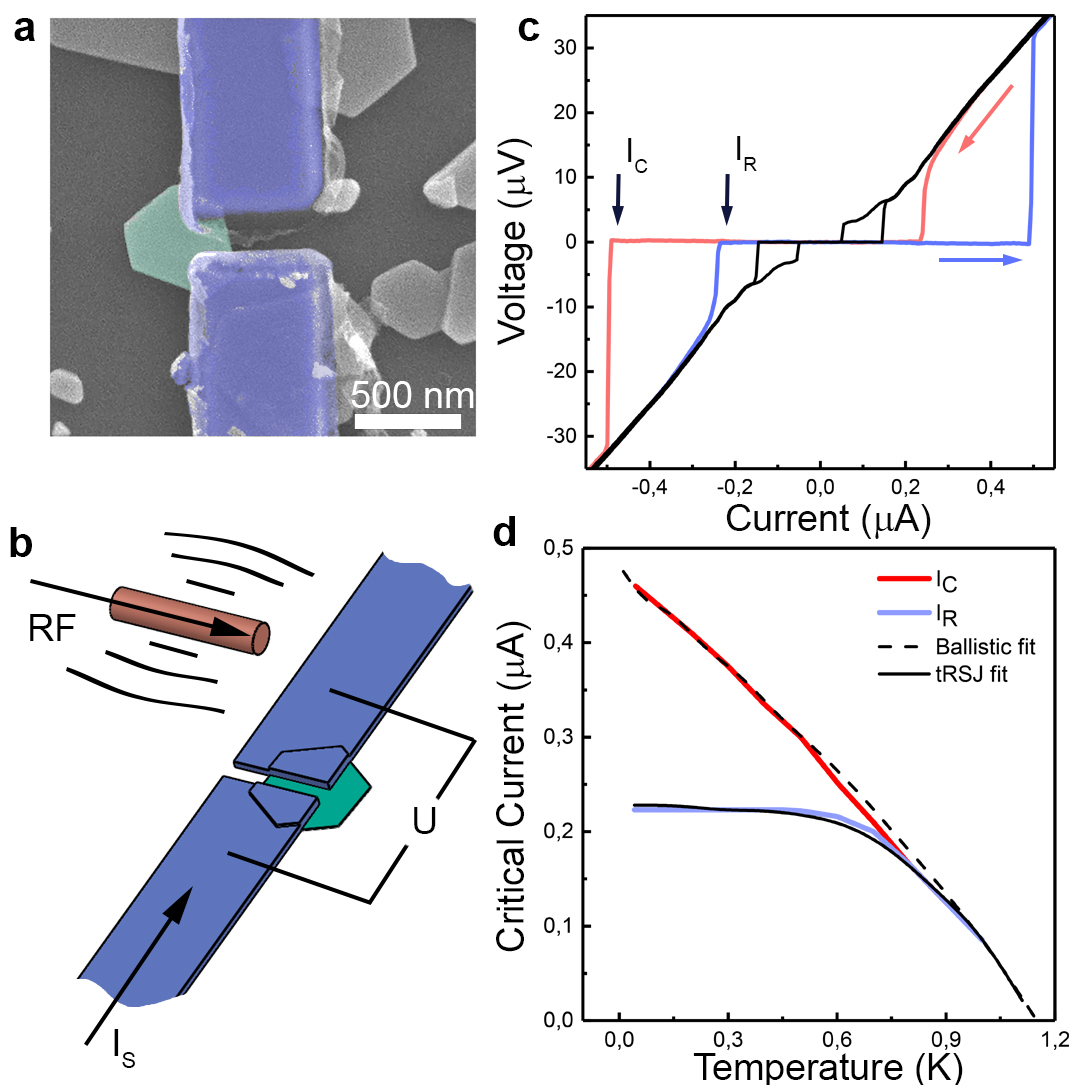}
\caption{\textbf{Experimental observation of supercurrent in Josephson junction based on ultra small single crystal of Bi$_2$Te$_{2.3}$Se$_{0.7}$.} \textbf{a} 
Scanning electron microscopy image of the Nb/Bi$_2$Te$_{2.3}$Se$_{0.7}$/Nb Josephson junction. The superconducting Nb leads (S-regions) are shown in blue color; the flake of Bi$_2$Te$_{2.3}$Se$_{0.7}$ in the N-region is colored light green. \textbf{b} Schematic view of the device along with the external electrical circuit. DC bias current $I$ is applied and the voltage drop $V$ across the junction is measured.  \textbf{c} Hysteretic $V(I)$ characteristic of the junction measured at $T=20\,$mK. Supercurrent at $V_{RF}=0$~$\mu$V is observed with a critical current $I_C=0.46\,\mu$A and a retrapping current $I_R=0.22\,\mu$A, respectively. The solid black line corresponds to the hysteretic $V(I)$ curve measured at $f_{RF}=1.5\,$GHz and $V_{RF}=20$~$\mu$V with Shapiro steps. \textbf{d} Critical $I_C$ and retrapping $I_R$ currents as function of $T$. The ballistic fit was performed using the Eilenberger equations. The retrapping current was fitted by t-RSJ model (see text for details).}
\label{fig:Fig.1} 
\end{figure}

Distinct features of $T$-ABS originate  from that fact that   a pair of Majorana real fermions    correspond  to a complex fermion  level. This level defines fermion parity $\sigma$ in the junction where, let us say,   $\sigma=+1$ if the level is occupied and $\sigma=-1$ if it is empty. Energies of $t$-ABS, given by $\epsilon_\sigma(\varphi)=\sigma E(\varphi)$, are non-degenerate with respect to  $\sigma$  and  the function $E(\varphi)$ is $4\pi$-periodic and odd. Unlike ABS in trivial SNS junctions,  the  branches $\epsilon_\sigma(\varphi)$ and $\epsilon_{-\sigma}(\varphi)$ have topologically protected crossings at zero energy in the phases of $\varphi=\pi+2\pi n, n \in  \mathbb{Z}$ (the ground state is degenerate at these points with respect to $\sigma$). Due to these protected crossings  the current-phase relationship    acquires anomalous $4\pi$-component, in addition to the conventional $2\pi$-periodic one. The sign of the $4\pi$-component can be positive or negative depending on the sign of the parity $\sigma$. 

The $4\pi$-component can be observed experimentally only if the parity does not change over the duration of the measurement. Otherwise, if the parity changes frequently, the $4\pi$-component contribution is averaged to zero. This explains why rapid dynamics measurements are preferable. Thus, previously, the $4\pi$-Josephson effect was probed in the high-frequency regime in which $\varphi$ grows in time and crosses the degeneracy points faster than a parity relaxation occurs \cite{rosen2021fractional}.

   The initial experimental confirmation of the absence of the first Shapiro step at a voltage $V_{\rm 1}=\frac{h f}{2 e}$ (where $h$ represents Planck's constant, $e$ the electron charge, and $f$ indicating frequency) was demonstrated in spin-orbit coupled nanowires \cite{rokhinson2012fractional, mourik2012signatures, das2012zero, Iorio2022Half}, as well as in three-dimensional topological insulators \cite{wiedenmann20164}. Later, HgTe/CdTe-based junction experiments unveiled the absence of the first nine odd steps \cite{bocquillon2017gapless} and the fractional Josephson radiation frequency. The debate surrounds the anomalous interpretation of the ac Josephson effect due to unbroken time-reversal symmetry, challenging expectations for localized Majorana states in Ref.~\cite{bocquillon2017gapless}.

Absence of $n=\pm 1$ Shapiro step was shown in exfoliated topological insulator Bi$_2$Se$_3$~\cite{le2019joule} flakes and Dirac semimetals \cite{li20184pi, rosenbach2021reappearance}. In particular, high-intensity drives exhibited residual supercurrent~\cite{le2019joule}, confirming the $4\pi$-periodic nature of the ac Josephson effect. The partial fractional observation of the Josephson effect linked to Joule overheating, the decrease in parity lifetime, was evident in hysteretic $V(I)$-curves. 

To explain the Missing First Step (MFS) effect, the phase winding frequency, $f_{Jt}=eV/h$, is assumed to be a half of the usual one, $f_{J}=2eV/h$, being a consequence of coherent $1e$ transport. Therefore, the first microwave-induced step should occur at a voltage $eV_{1t}/h=hf/e$, for a given radiation frequency $f$, which is twice higher than the usual voltage for the first step,$V_1=hf/2e$. Thus, it appears that the first step is missing, which is the MFS effect ~\cite{le2019joule, houzet2013dynamics, jiang2011unconventional, wiedenmann20164}.  An observation of the   absence of odd Shapiro steps   as well as fractional frequency  of  Josephson radiation   is believed to be one of possible routes to prove the existence of the Majorana zero modes ~\cite{dominguez2012dynamical, badiane2013ac, virtanen2013microwave}.

\begin{figure*}[t!]
\centering
\includegraphics[width=17cm]{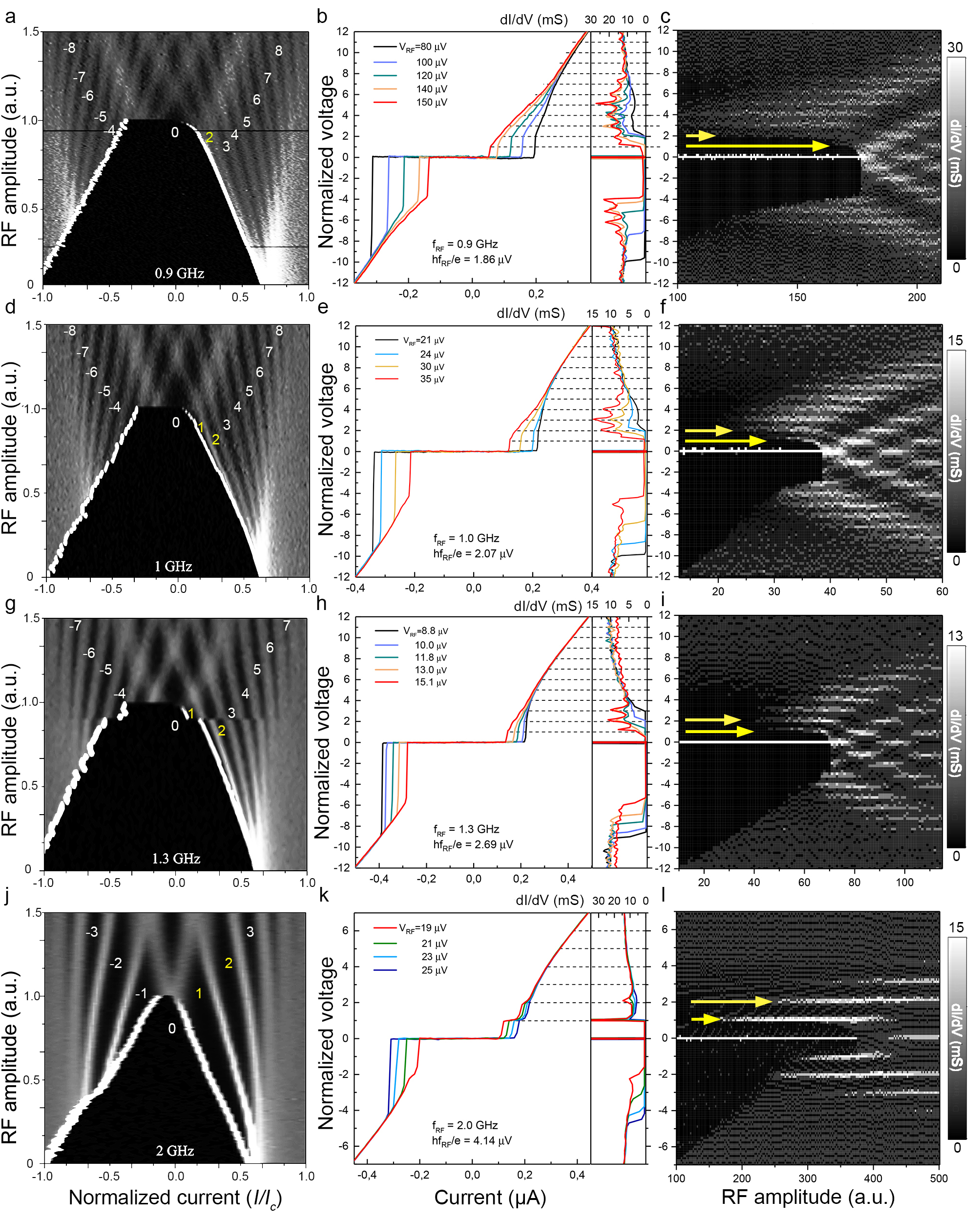}
\caption{\label{fig:Fig. 2}\textbf{Shapiro maps dV/dI for different RF drive frequencies.} \textbf{a, d, g} and \textbf{j} The drive frequencies $f_{RF}=$ 0.9, 1, 1.3, and 2\,GHz are plotted. The maps cover negative (switching) and positive (retrapping) polarities of the bias current and demonstrate a hysteresis. The left-hand side of a given map for $I<0$ corresponds to the switching current, and the I$>$0 segment corresponds to the retrapping current. \textbf{b, e, h} and \textbf{k} $V(I)$-curves in the normalized unit for voltage. The negative polarity shows the switching current, and the positive polarity shows the retrapping current. For I$>$0, the even-odd effect is observed. The panel on the right shows the dI/dV as a function of bias current at different RF powers, giving a clearer view of the effect. The second Shapiro step emerges at a lower radiation power than first step. The first step is recovered at sufficiently large RF drive amplitudes.}
\end{figure*}
      
   Comparing to trivial SNS junctions, ballistic SNS junctions exhibit ABS energies with a $4\pi$-periodic dependence \cite{galaktionov2021fractional} $\pm\Delta\cos(\phi/2)$, where $(\phi/2)$, is the superconducting phase difference and $\Delta$ is the induced energy gap. However, recent theory indicates a $2\pi$-periodic static CPR \cite{galaktionov2021fractional} in this case.

The MFS effect could also originate from trivial ballistic junctions, posing a challenge to differentiate between possibilities. The ubiquity of the MFS effect includes observations in amorphous superconducting nanowires and nanobridges ~\cite{bae2009current}. First steps or few steps could be masked by junction switching or retrapping currents. Junction bistability, characterized by abrupt voltage jumps, can hinder first step visibility. Recovery of the first step involves increasing microwave frequency or power.

Distinguishing topological and trivial effects involves a proposed method~\cite{le2019joule} involving quasiparticle poisoning due to Joule heating. Quasiparticles cause rapid changes in the Majorana Zero-Mode (MZM) contribution, averaging it to zero. This explains why only the first step disappears while the third step remains, as Joule heating creates high quasiparticle population. High-population quasiparticles can occupy Majorana fermion modes, altering parity according to theory~\cite{beenakker2014annihilation}. Thus, changing current from high to zero should retain the presence of the first step.

 In this letter, we demonstrate the anomalous Josephson effect, namely the absence of the first Shapiro step. The studied SNS junction involves two Nb electrodes coupled through an ultrasmall single nanocrystallite of a 3D topological insulator Bi$_2$Te$_{2.3}$Se$_{0.7}$ ~\cite{stolyarov2020josephson}, acting as the normal (N) region of the SNS jucntion. 
   We present Shapiro step maps as the differential conductance $d I/dV$ plotted versus the DC current bias and the RF drive power, in the 0.5 to 3 GHz frequency range. Strongly hysteretic curves $V(I)$ and the $I_C$(T) dependence are also presented. The $I_C$(T) function is well described by the ballistic electron theory of the SNS junctions (see also ref\cite{stolyarov2020josephson}). To describe the Shapiro steps and their dependence of the RF drive power, the two-channel thermal resistively shunted junction (t-RSJ) model is adopted. Our important conclusion is that the first step is missing even in the retrapping branch of the $V(I)$ curve, i.e., when the Joule heating is quite significant. The model we use to analyze the data includes (i) quasiparticle overheating due to the Joule effect \cite{le2019joule}, and (ii) thermally activated poisoning of the MBS. We demonstrate a good agreement of the results with this extended t-RSJ model.

\section{\label{sec:Experiment} Experiment}

The Bi$_2$Te$_{2.3}$Se$_{0.7}$ nanocrystalline ingots used to make our junctions were synthesized by Physical Vapor Deposition (PVD) method ~\cite{yakovlev2022physical}. Our detailed XRD analysis shows that the ingots contain only a single phase of Bi$_2$Te$_{2.3}$Se$_{0.7}$.The studied SNS junction is shown in  Fig.\ref{fig:Fig.1}~(a). The measured thickness of the flake is $t=20$ nm. The length of the N-region is $L=150$, nm, and the widths of the Nb electrodes are $w=500$, nm. Several junctions of different sizes were fabricated and studied in the  experiments. 
Fig.\ref{fig:Fig.1} (b) shows a schematic view of the Josephson junction and the external electrical circuit is presented. The device is measured in a regime of DC bias current $I$ where the voltage drop $V$ across the junction is present; the RF drive is applied through a microwave generator.  

The normal state resistance of the junction at room temperature is $R_N\approx 1.2$ k$\Omega$. The resistance shows a conventional  metallic  behavior if the temperature $T$ is decreased. An abrupt drop in resistance was detected at $T=8$, K attributed to the superconducting phase transition in the Nb electrodes that form the junction. The further cooling down of the junction shows a smooth decrease of the resistance in a domain from 5 to 1\,K. The latter is characteristic for the superconducting proximity  effect induced in the TI flake by the  closely spaced Nb electrodes \cite{stolyarov2020josephson}.

Fig. \ref{fig:Fig.1}(c) illustrates a hysteric voltage-current $V(I)$ relation, where the critical current ($I_C$) and the retrapping current ($I_R$) are not equal. The red curve represents the increasing bias current and shows the switching (critical) current, a jump-like transition from superconducting to normal state. The blue curve corresponds to the opposite direction of the current ($I$) and represents the retrapping current. The black curve illustrates the voltage-current characteristic under the influence of an RF signal.
Fig. \ref{fig:Fig.1}(d) displays the temperature dependence of both critical and retrapping currents. Hysteresis appears at temperatures below 0.7 K. Additionally, the absence of a low-temperature saturation in $T_c(T)$ suggests ballistic transport \cite{stolyarov2020josephson}, which is further supported by fits generated using the ballistic-limit Eilenberger equations and their solutions obtained by Galaktionov and Zaikin \cite{galaktionov2021fractional} . 

The presence of an RF antenna \ref{fig:Fig.1}(b), which is not ideally matched and is located close to the sample, leads to an elevation in background noise levels. This effect arises due to the direct connection of the system to the room-temperature electronics. In contrast to our previous work, in the study referenced in Ref.\cite{stolyarov2022resonant}, where the contribution of s-wave and p-wave superconductivity was investigated. The critical current didn't achieved higher magnitudes related to p-wave superconductivity.  In previous work the sample was meticulously shielded from all forms of electromagnetic radiation within the cavity. Additionally, it featured supplementary low-pass filters integrated with silver paste. Nevertheless, the critical current remains large (at s-wave level) \ref{fig:Fig.1}(d), and the transport mode remains predominantly ballistic, even in the presence of an additional noise source like the antenna.

 The hysteretic behavior of the $V(I)$-curves indicates Joule overheating effects for quasiparticles \cite{courtois2008origin}. This is considered a source of quasi-particle poisoning \cite{karzig2021quasiparticle}. In addition, we discuss the effect of microwave radiation in detail. In Figs.\ref{fig:Fig. 2} (a,b,c) we show the results obtained at $f_\textit{RF}=0.9$\,GHz. Five representative $V(I)$-curves, corresponding to different amplitudes of the RF radiation, are shown in Fig.\ref{fig:Fig. 2} (b); the insert shows the $dV/dI$, which amplifies the steps of the $V(I)$-curve. The corresponding map of the differential conductance, plotted versus the bias current and the RF signal amplitude, is shown in Fig.\ref{fig:Fig. 2}(a). The darker regions represent voltage plateaus corresponding to a lock-in synchronization effect between the RF field and the evolution of the superconducting phase. In other words, the darker regions are Shapiro steps, which are numbered in Fig.\ref{fig:Fig. 2}(a). 
 
 A prominent region is the large black central region which represents the perfect superconductivity (zero voltage) of the sample. In this zero resistance region (0-region) no steps can be observed. 
 On the left side, this superconducting region is bounded by the critical current dependent on the RF amplitude $I_C(V_{RF})$. On the right side the 0-region is bounded by the retrapping current which is also dependent on the RF amplitude $I_R(V_{RF})$. The 0-region, if it overlaps some of the Shapiro steps, can completely suppress them since the voltage is exactly zero in the 0-region. 
 
 For example, on the left side, and at low bias, the first visible step is number 10, that is, the first through the ninth steps are suppressed (see Fig.\ref{fig:Fig. 2} (b-insert)). The dependence of the critical current on the RF amplitude and the dependence of the steps (bright regions) on the RF amplitude are both linear, but the slope of the critical current is larger. Therefore more step boundaries cross the critical field line at higher driving amplitudes. Thus, at higher RF amplitudes, it is possible to observe lower-order steps. For example, in the critical current branch, at the lowest power the lowest step is $n=10$, while at the highest power the lowest step is $n=4$ (Fig.\ref{fig:Fig. 2}(a)). 
 
 The situation with the retrapping current is more favorable for the observation of the lower-order steps. This is simply because $I_R<I_C$, so the voltage explores lower values on the retrapping branch of the $V(I)$-curve (positive region in Fig.\ref{fig:Fig. 2} (a)), i.e., on the branch on which the current is ramped down. Because of this, all steps are observed on the retrapping branch, except the first one. 
 
 The general condition for the observation of a step number $n$ is: $V(I_C)<V_n=nhf/(2e)$ on the critical current branch and $V(I_R)<V_n=nhf/(2e)$ on the retrapping branch. On the other hand, as the amplitude of the RF signal increases, both the critical current and the retrapping currents are reduced. Thus the number of steps which can be detected increases with the amplitude of the RF signal.
 
 Another map of differential conductance is shown in Fig.\ref{fig:Fig. 2}(c), which represents $dV/dI$ plotted versus the RF amplitude and the DC voltage on the sample, $V$, normalized by the fundamental mode voltage of the phase evolution, $hf/2e$. Interestingly, it shows that the second step is present at very low amplitudes. However, the first step appears only when the driving amplitude is quite large and approaches 170 $\mu$V.
 
 When the critical current branch is consider, up to 9 first steps appear suppressed. This suppression is most probably due to the switching-current voltage jump being larger than $nhf/2e$. Thus, in what follows, we focus on the retrapping branch, where only the first step is absent, while the second is observed.There could be multiple explanations for this phenomenon of the first step disappearance. It could be due to the topological nature of the proximitized material in these sample. As noted previously, Majorana fermions support coherent transfer of single electrons and the corresponding CPR is proportional to $\sin(\phi/2)$, where $\phi$ is the phase difference at the junction. Such a CPR could lead to the disappearance of the first Shapiro step. Another possibility is that the ballistic character of the electron transport through the junction causes the second step to become more prominent than the first one. According to the recent Galaktionov-Zaikin model \cite{galaktionov2021fractional}, the second step could occur at a lower driving power than the second step if the phenomenon of MAR is taken into account. Yet another possible explanation is that the second step boundary crosses the curve $I_R(V_{RF})$ at a lower $V_{RF}$, where $V_{RF}$ is the amplitude of the RF driving signal. This seems quite plausible based on the results shown on Fig.\ref{fig:Fig. 2}(a).  
 
 Considering results obtained at higher frequencies, we focus on the retrapping branch (positive current bias on all plots), since the switching current branch produces a large jump at the critical current such that the voltage immediately goes to value much higher than needed for the observation of low-order steps. Yet, it is interesting to note that, at $f=2\,\mathrm{GHz}$, even on the critical current branch, the first observed step is $n=2$, while at a higher RF drive the $n=1$ is also visible (Fig.\ref{fig:Fig. 2}(j)). 
 
 \begin{figure}[t!]
\centering
\includegraphics[width= 8.5 cm]{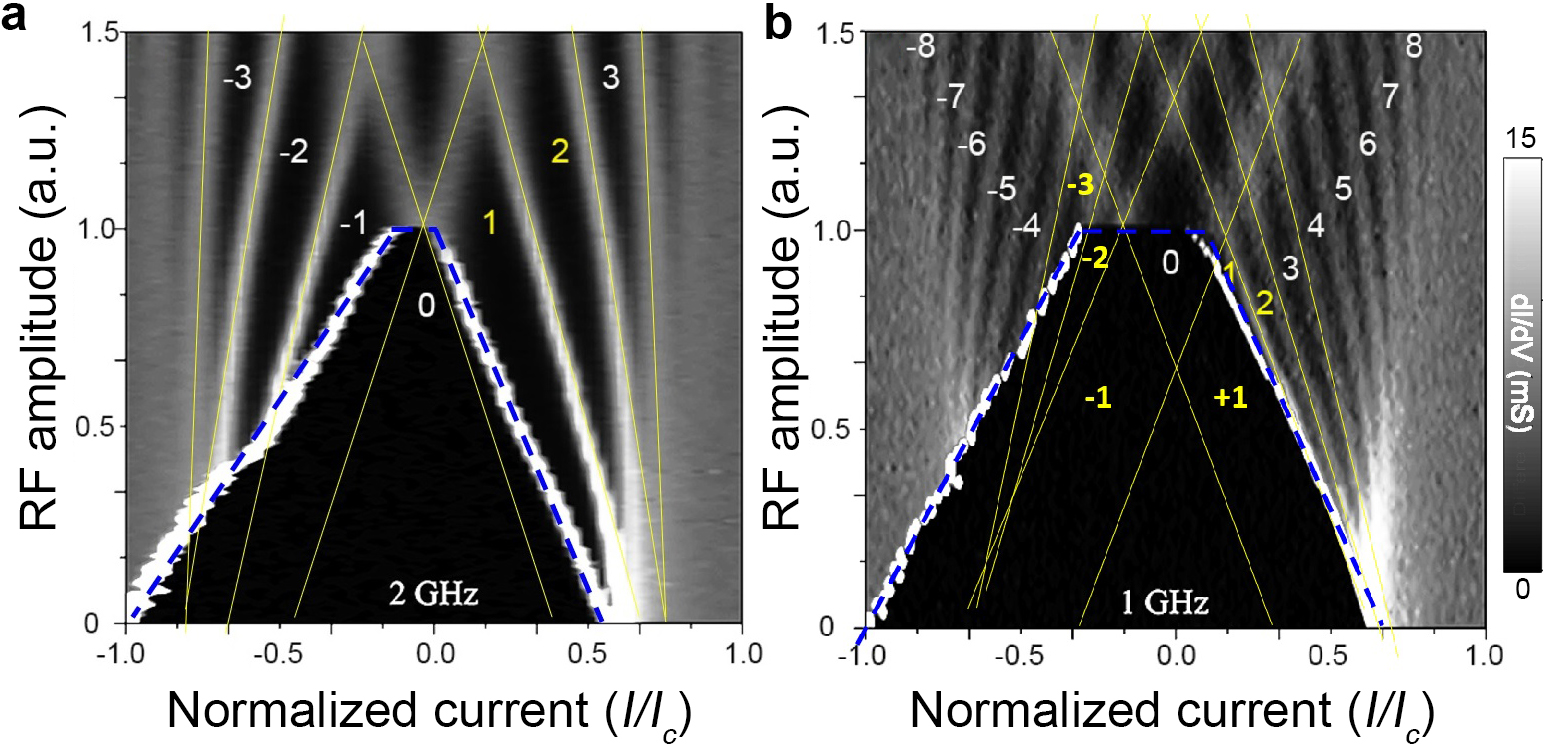}
\caption{\label{fig:Fig.3} \textbf{Retrapping current overlapping Shapiro steps.} \textbf{a} and \textbf{b} Differential resistance versus normalized bias current and RF drive amplitude. The retrapping currents and the switching currents (blue dashed lines) form the main borders of a trapezoidal zero-resistance region. Microwave-induced steps are marked by yellow lines. (\textbf{a}) $f=2$\,GHz, The first step ($V=hf/2e$) is observable at low RF drive. (\textbf{b}) $f=1$\,GHz. The first step is not always apparent. At low RF drive the junction transitions from zero resistance to the second step ($V=hf/e$), without entering the first step.}
\end{figure}
 
 Let us now return to the discussion regarding the retrapping branch. In this case, we will focus on the data acquired at 1\,GHz, as depicted in Fig.\ref{fig:Fig. 2}(d), (e), and (f), following the same format as previously discussed. Similarly to the results of 0.9\,GHz, the second step at this frequency appears at a lower RF amplitude compared to the first step, as shown in Fig.\ref{fig:Fig. 2}(f). This outcome is an exact analogy to the behavior observed at 0.9\,GHz.

However, the behavior of the sample changes noticeably when it moves to 1.3\,GHz, as illustrated in Figs.\ref{fig:Fig. 2}(g), (h), and (i). At this frequency, no steps are observed in the retrapping current branch at the lowest $V_{RF}$ value. As the driving amplitude increases, the first and second steps become visible approximately at the same level of the RF drive, as depicted in Fig. \ref{fig:Fig. 2}(i). This behavior aligns with the general expectation that an increase in the frequency of the RF drive leads to larger step sizes ($hf/2e$) capable of reaching the voltage level on the sample just before the retrapping event occurs and causes the voltage to drop to zero.

Finally, the data acquired at 2\,GHz is presented in Figs. \ref{fig:Fig. 2}(j), (k), and (l). At this frequency, the multiple fractional Shapiro steps effect is not observed. As the current increases, the first step appears at a voltage of $V=hf/2e$, followed by another step at a higher current corresponding to $V=hf/e$. In other words, both the first and second steps are observed. This qualitatively distinct behavior arises from the fact that the slope of the $I_R(V_{RF})$ line is now lower than the slope of the first step, whereas at 1.3\,GHz, the slope of the first step line was lower than the slope of the retrapping current line. 

\section{\label{sec:Discussion}Discussion}

The studied SNS junction involves two Nb electrodes coupled through an ultrasmall single nanocrystallite of a 3D topological insulator Bi$_2$Te$_{2.3}$Se$_{0.7}$ ~\cite{stolyarov2020josephson}, acting as the normal (N) region of the SNS junction. 
   We present Shapiro step maps as differential conductance $d I/dV$ plotted versus DC current bias and RF drive power, in the frequency range from 0.9\,GHz to 3\,GHz. Strongly hysteretic I(V) curves as well as the $I_C$(T) dependence are also shown. The $I_C$(T) function is well described by the ballistic electron theory of SNS junctions (see also ref\cite{stolyarov2020josephson}). In such measurements the Joule heating is always present, down to the retrapping occurrence. Thus, the expectation is that the parity of the MZM (if MZM is indeed present) should be changing very rapidly and so the first step should be observed. Our important conclusion is that the first step is still missing in the retrapping current branch, i.e., a strong Joule heating, which is responsible for a pronounced hysteresis of the V-I curves, is not able to eliminate the MFS effect. We also analyze the critical current as a function of temperature and demonstrate that our SNS junctions are ballistic, because of the high structural perfection of the PVD-grown topological insulator crystals. Thus, our results provide evidence that the incomplete even-odd effect (the MFS effect) can occur in ballistic junctions even under strong quasiparticle poisoning. We compare the results to the models outlined above.  

A general feature, observed at all frequencies, is that all steps are visible if the RF signal is sufficiently strong (see the top parts of the plots in Figs.\ref{fig:Fig.3}(a,b)). The reason is that at sufficiently large $V_{RF}$ the critical current and the retrapping current are both zero, while the superconducting order parameter is not suppressed to zero. Under such conditions, the DC voltage increases gradually as the DC bias current increases. Thus, all low-order steps are visible and there are no missing steps at sufficiently large amplitudes of the RF drive, $V_{RF}$. In Figs.\ref{fig:Fig.3}(a,b) we mark all steps with yellow lines; the switching and retrapping currents are shown by blue dashed lines. The main difference between the high-frequency data (2\,GHz) (Fig.\ref{fig:Fig.3}(a)) and the low-frequency data (1\,GHz) (Fig.\ref{fig:Fig.3}(b)) is that in the first case the slope of the yellow lines is larger than the slope of the retrapping current blue dashed lines, while in the latter case the slopes are about equal. Also, the spacing between the yellow lines (the steps) is naturally lower if the RF signal has a lower frequency, since the voltage difference between the steps is $hf/2e$, where $f$ is the frequency of the RF signal. On the other hand, at a high RF frequency, there is a significant region between the retrapping line (blue) and the second yellow line. Thus, a part of the first voltage plateau can be observed experimentally even at low RF drive (Figs.\ref{fig:Fig.3}(a,b)).

An alternative and more traditional explanation of the observed results is given below. The hysteretic behaviour is linked to the Joule overheating effect related to the phase winding and the corresponding voltage. This is considered to be a source of quasiparticle poisoning. The poisoning results in the partial observation of $4\pi$-periodic Josephson effect where only $n=\pm 1$ Shapiro steps are absent. Our results differ from those previously reported in that we observe the absence of the first step in the retrapping current, when the quasiparticle poisoning is not negligible.
 
\begin{figure}[t!]
\centering
\includegraphics[width=8.5 cm]{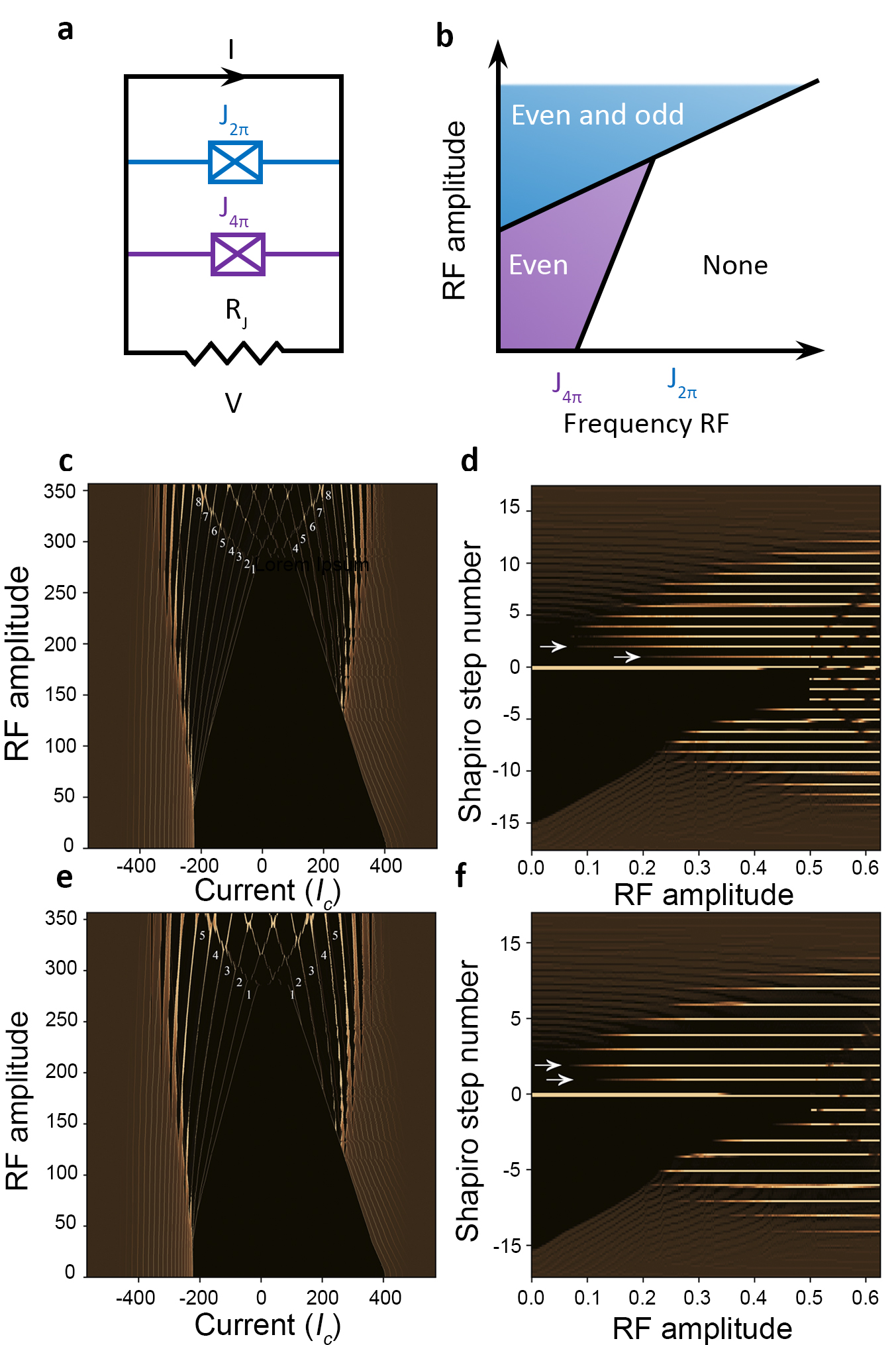}
\caption{\label{scheme} \textbf{a} Scheme of the two-channel thermal RSJ model. One of the two parallel Josephson junctions (J$_1$) stands for the trivial supercurrent and the other one (J$_2$) stands for the topological channel in parallel with shunting resistance $R_n$. \textbf{b} Phase diagram of odd and even Shapiro steps in two-channel thermal RSJ model. Simulated differential resistance map as a function of the RF current amplitude for RF current frequencies (\textbf{c}) 2 GHz and (\textbf{e}) 0.9 GHz with 2$\pi$ and 5\% 4$\pi$-component. Corresponding Shapiro bins maps for RF current frequencies (\textbf{d}) 2 GHz and (\textbf{f}) 0.9 GHz.}
\end{figure}

We model our system using the two-channel thermal RSJ model proposed in ref.\cite{le2019joule, de2016interplay}. This model consists of two parts: additional coherent superconducting channel originating from Majorana bound states and contributing $\sin(\phi/2)$ supercurrent, and self-consistent thermal balance which is manifested in the hysteresis of the $V(I)$ curves. In Fig.\ref{scheme}(a) the scheme of the two-channel thermal RSJ model is shown. There are two parallel superconducting junctions in parallel with the shunt resistance $R$. The Josephson junction J$_1$ is a conventional junction with $2\pi$-CPR, and the second junction J$_2$ is a topological one with $4\pi$-CPR. 

According to this total superconducting current is $I_{s}(\varphi) = I^{2\pi}_{c}\sin(\varphi)+I^{4\pi}_{c}\sin(\varphi/2)$ and equation for phase can be written as
\begin{equation}
\frac{\hbar d \varphi}{2eR d t}=I_{DC}+I_{RF} \sin \left(\omega_{RF} t \right)-\left(I^{2\pi}_{c} \sin (\varphi)+I^{4\pi}_{c} \sin (\varphi/2)\right),
\label{RSJ}
\end{equation}
where I$_{DC}$ is the DC bias current, I$_{RF}$ is the RF current applied by a generator, $\omega_{RF}$ is the frequency of RF current, $I^{2\pi}_{c} \left(I^{4\pi}_{c}\right)$ is the critical current of $2\pi$ $(4\pi)$ component, $R$ is the normal state resistance of the junction.

To take into account the hysteretic behavior of $V(I)$ characteristics due to Joule overheating, we write the heat balance equation:
\begin{equation}
\langle P(t)\rangle=\Sigma U\left(T_{e}^{5}-T_{ph}^{5}\right),
\end{equation}
where $\Sigma$ is the electron-phonon coupling constant of the normal material, $U$ is the effective volume of the sample, $T_{e}$ and $T_{ph}$ are electron and phonon temperatures respectively. A self-consistent scheme is provided as follows: at a given I$_{DC}$, I$_{RF}$ we begin by supposing a temperature $T$ (at the first iteration $T=T_{ph}$ is equal to the bath temperature of the cryostat) which gives a critical current $I_{c}(T) = I^{2\pi}_{c}(T) + I^{4\pi}_{c}(T)$. Next we solve equation \ref{RSJ} from which we estimate the Joule power $P = \langle I(t)V(t)\rangle$ that gives us temperature of electron subsystem $T_e=\sqrt[5]{{T_{ph}}^{5}+\frac{P}{\Sigma U}}$ and hence a new critical current $I_{c}(T_{e})$. Then this sequence is repeated until the difference between a new estimated temperature $T^{n+1}$ and that from the previous step $T^{n}$ is small enough. The temperature dependence of the critical current $I_{c}(T)$ is taken from the ballistic fit and values of parameters $\Sigma$ and $U$ are extracted from the fit of retrapping $I_{r}$ current (see Figs.\ref{fig:Fig.1}).

As it has been shown in the work by Le Calvez et al. \cite{le2019joule} there are two characteristic frequencies in this system $f^{2\pi}=\frac{1}{\hbar}2eRI^{2\pi}_{c}$ and $f^{4\pi}=\frac{1}{\hbar}eRI^{4\pi}_{c}$. They define the 2D phase diagram, see Fig.\ref{scheme}(b).In Fig. \ref{scheme}(c,e), a numerical calculation was performed using the model for a frequency of 0.9\,GHz. The results demonstrate that at low power levels of the RF signal, the second stage appears first, while at high power levels, the first stage appears. To achieve the closest resemblance to the experimental data, it was necessary to include a minimum of 5\% of the $4\pi$-periodic component. Fig.\ref{scheme}(e,f) depicts the fitting results for a frequency of 2\,GHz, which replicate the observed behavior in the experimental data. An interesting result of the model is that it reproduces the zero-resistance region at low DC and RF bias and predicts a linear shape of the boundary. 

Our study sheds light on the perplexing phenomenon of selective missing of the first Shapiro step exclusively in topological Josephson junctions. In our ballistic Josephson junction, constructed using the 3D topological insulator Bi$_2$Te$_{2.3}$Se$_{0.7}$, this phenomenon is aptly characterized and, can be effectively explained by a two-channel thermal RSJ model.


\begin{acknowledgments}

\end{acknowledgments}

\end{document}